\documentclass[12pt]{article}
\usepackage{graphicx}
\usepackage{dcolumn}
\usepackage{bm}
\usepackage{graphics}
\usepackage{amsmath}
\usepackage{amssymb}
\usepackage{amscd}
\usepackage{afterpage}
\usepackage{float,times}
\usepackage{subfigure}
\usepackage{rotating}
\usepackage{multirow}
\usepackage{fancyheadings}
\usepackage{epsfig}
\usepackage{theorem}
\usepackage{moreverb}
\usepackage{euscript}
\usepackage{psfrag}
\DeclareMathOperator{\sgn}{sgn}
\begin{document}

\begin{center}
{\hbox to\hsize{\hfill June 2007 }}

\bigskip
\vspace{3\baselineskip}

{\Large \bf

Particle interference as a test of Lorentz-violating electrodynamics \\
}
\bigskip

\bigskip

{\bf 
Archil Kobakhidze and Bruce H. J. McKellar \\}
\smallskip

{ \small \it School of Physics, Research Centre for High Energy Physics, \\ The University of Melbourne, Victoria 3010, Australia\\}

\bigskip

\vspace*{.5cm}

{\bf Abstract}\\
\end{center}
\noindent
{\small 
In Lorentz-violating electrodynamics a steady current (and similarly a static charge) generates  both static magnetic and electric fields. These induced fields, acting on interfering particles, change the interference pattern.  We find that   particle interference experiments are  sensitive to small Lorentz violating effects, and thus they can be used to improve current bounds on some Lorentz-violating parameters.}  

\bigskip

\bigskip

\bigskip



\paragraph{Lorentz-violating electrodynamics.} 
Historically the theory of Special Relativity has been established by studying properties of light which are described by Maxwell's electrodynamics.  Special Relativity  identifies the Poincar\'e group as a spacetime symmetry group that underlies all fundamental physical laws of nature. Although no decisive departure from relativistic invariance has been observed so far, theoretical and experimental studies of possible small violations of this fundamental symmetry continue to attract considerable attention (see \cite{Kostelecky:2005mj} and references therein). 

Currently the best laboratory tests of relativistic invariance are modern versions of the Michelson-Morley and  Kennedy-Thorndike experiments using resonant cavities \cite{Wolf:2004gg}. In this Letter we propose a new class of experiments to search for possible  violation of Lorentz invariance  in electrodynamics. Namely, we demonstrate that the standard quantum-mechanical particle interference can be significantly altered due to the effects which are entirely attributed to the Lorentz-violating interactions, and thus particle interference experiments can be used to improve current bounds on some Lorentz-violating parameters.

Let us consider  Lorentz-violating model of electrodynamics described by the Lagrangian \cite{Kostelecky:2002hh},
\begin{eqnarray}
{\cal L}=-\frac{1}{4}F_{\mu\nu}F^{\mu\nu}-j_{\mu}A^{\mu} \nonumber \\
-\frac{1}{4}(\kappa_
F)_{\mu\nu\rho\sigma}F^{\mu\nu}F^{\rho\sigma}+\frac{1}{2}(\kappa_
{AF})^{\alpha}\epsilon_{\alpha\mu\nu\rho}A^{\mu}F^{\nu\rho}
\label{1}
\end{eqnarray}
The 23 constant parameters $(\kappa_
F)_{\mu\nu\rho\sigma}$ and $(\kappa_
{AF})^{\mu}$ define strength of the  Lorentz-violating interactions in the photon sector. The matter sector is assumed to be Lorentz invariant\footnote{Actually some of the parameters can be moved from the photon sector to the matter sector by a suitable coordinate transformations \cite{Bailey:2004na}
. Since the theory is invariant under the coordinate (passive) Lorentz transformations the physical effects  must be the same in both coordinate frames.}.  Consequently, the ordinary Lorentz-force law is intact. It is convenient to introduce the following notations:
\begin{eqnarray}
(\kappa_
{DE})^{ij}=-2(\kappa_
F)^{0i0j}~, \nonumber \\
(\kappa_
{HB})^{ij}=\frac{1}{2}\epsilon^{ipq}\epsilon^{jrs}(\kappa_
F)^{pqrs}~, \nonumber \\
(\kappa_
{DB})^{ij}=-(\kappa_
{HE})^{ji}=(\kappa_
F)^{0ipq}\epsilon^{jpq}~. 
\label{2}
\end{eqnarray} 
We further simplify the model by setting to zero those of the above parameters which are known to be strongly  constrained  from observations.  Non-observation of birefringence effects in the light propagating from distant astrophysical sources is consistent with $(\kappa_
{AF}) \lesssim10^{-42}$ \cite{Carroll:1989vb}
 and  $(\kappa_
{DE}+\kappa_
{HB})$, $(\kappa_
{DB}-\kappa_
{HE}) \lesssim 10^{-37}$ \cite{Kostelecky:2002hh}, \cite{Kostelecky:2006ta}.  We simply take $\kappa
_{AF}=0$,  and $\kappa
_{DE}=\kappa_
{HB}=0$. As $\kappa
_{DB}=\kappa_
{HE}$ is an antisymmetric $3\times 3$ matrix,  we are left only with 3 parameters, $\kappa^{i}=\frac{1}{2}\epsilon^{ijk}(\kappa_
{DB})^{jk}$. These parameters are currently relatively less constrained. The bounds $\kappa^i \lesssim 10^{-11}$ have been obtained in experiments with optical and microwave cavities \cite{Wolf:2004gg}. 

With the above simplifications the modified source dependent Maxwell's equations become:
\begin{equation}
\vec {\triangledown}\cdot \vec{E} + \vec{\kappa}\cdot \left(\vec{\triangledown} \times \vec{B}\right)= \rho~,  
\label{3}
\end{equation}
\begin{equation}
\vec{\triangledown}\times \vec{B}-\frac{\partial \vec{E}}{\partial t}+\vec{\triangledown}\times\left(\vec{E}\times \vec{\kappa}\right)-\frac{\partial \left(\vec{B}\times \vec{\kappa}\right)}{\partial t}=\vec{j}~,
\label{4}
\end{equation}
The remaining homogeneous equations stay the same: $\vec{\triangledown}\cdot \vec{B}=0$, $\vec{\triangledown}\times \vec{E}+\frac{\partial \vec{B}}{\partial t}=0$.  From the above equations we see that even in the static case the magnetic and electric fields do not decouple in  Lorentz-violating electrodynamics.  A steady current (and similarly a static charge) generates  both static magnetic and electric fields. Thus if for example, an electron is moving near the static magnetic  source it will experience a Lorentz force due to the induced static electric field. This effect can be detected in electron interference experiments.   In what follows  we estimate the sensitivity of such interference experiments to Lorentz-violating parameters.

\paragraph{Lorentz-violating effects in particle interference experiments.}
Let us consider the typical interference set-up: a coherent beam of electrons is split into two parts, beam $A$ and beam $B$, each moving along $x$-direction ($y_A=a$ and $y_B=b$) of $xy$-plane.  
The beams produce a particle interference pattern on a screen located at a distance $L$ from the double slit. At the origin of $xy$-plane we place a long solenoid of radius $R$ ($R<|a|,|b|$) which carries a current density $\vec j=j_{\phi
}\vec\phi\delta(r-R)
$, where $\vec \phi
$ is an unit vector of the orthonormal basis in the cylindrical coordinates $(r, \phi
, z)$ ($\vec {X}=X_r\vec{r}+X_{\phi
}\vec{\phi
}+X_z\vec{z}$).  
In Maxwell's electrodynamics, the region where the beam propagates is force-free. In Lorentz-violating electrodynamics there is an induced classical electric field around solenoid which acts on  a particle in a  beam.  In the leading order in Lorentz-violating parameters the induced static electric field satisfies the equation (see eqs (\ref{3}) and (\ref{4})), 
\begin{equation}
\vec{\triangledown}\cdot \vec{E}=-\kappa B\sin(\alpha-\phi
)\delta(r-R)~,
\label{5}
\end{equation}
where $\kappa=\sqrt{\kappa_r^2+\kappa_{\phi}^2}$ and $\kappa_x=\kappa\cos\alpha,~k_y=\kappa\sin\alpha$. Although the magnetic field of the solenoid is also modified, the correction is of the second order in Lorentz-violating parameter, so we have taken  $\vec{B}=\vec{z}j_{\phi
}\theta(R-r)+{\cal O}(\kappa^2)$ in (\ref{5}) ($B=j_{\phi}$).  

We solve the equation (\ref{5}) in terms of the scalar potential $\Phi$ ($\vec{E}=-\vec{\triangledown}\Phi$, since $\vec{B}$ is static):
\begin{equation}
\Phi(r)=\sin(\alpha-\theta)B\kappa \left(\frac{R^2}{r}\theta(r-R)+r\theta(R-r)\right)~.
\label{5a}
\end{equation}

 Now, since the matter sector is undeformed, the effective Lagrangian for a non-relativistic point particle with the charge $e$ moving in the external electromagnetic field is unmodified as well,
 \begin{equation}
 L=\frac{p^2}{2m}-e\Phi + \frac{e}{m}\vec{A}\cdot \vec{p}~,  
 \label{6}
 \end{equation}  
where  $\vec{A}$ the vector potential, $\vec{B}=\vec{\triangledown}\times \vec{A}$. The Hamiltonian then reads:
\begin{equation}
H=\frac{1}{2m}(\vec{p}-e\vec{A})^2+e\Phi~.
\label{7}
\end{equation}

The scalar potential $\Phi$ (\ref{5a}) in equation (\ref{7})  induces and extra path-dependent phase which (in the WKB approximation) can be calculated straightforwardly:
\begin{eqnarray}
\delta=\delta_A-\delta_B=\int_{-l_1}^{l_2}\sqrt{p_0^2-2me\Phi(x,a)}dx -\int_{-l_1}^{l_2}\sqrt{p_0^2-2me\Phi(x,b)}dx\approx \nonumber \\
\frac{e}{v_0} \int_{-l_1}^{l_2}(\Phi(x,a)-\Phi(x,b))dx\approx
\frac{eBR^2}{v_0}\left[\kappa_y\log\left(\frac{l_1}{l_2}\right)- \sgn(a)\pi\kappa_x \right] - \left(a \to b\right)~.
\label{9}
\end{eqnarray}
In the above formulae $v_0$ is initial velocity of the electrons and, $e=0.302$, is the elementary charge in natural units. The distance between the double slit and solenoid is $l_2$, while $l_1$ is the distance between the solenoid and the screen, $L=l_1+l_2$. The last last step  of eq. (\ref{9}) we approximate using  $l_1, l_2>>|a|,|b|$.

Let us consider first the usual Aharonov-Bohm set up \cite{Aharonov:1959fk} where the solenoid is placed between the beams A and B (e.g, $a=-b$, being positive for definiteness). The presence of the vector potential in the first term of eq. (\ref{7}) generates the topological (path-independent) Aharonov-Bohm
 phase,
\begin{equation}
\delta_{\rm AB
}=e\oint\vec{A}\cdot d\vec{x}~.
\label{8}
\end{equation}
The Lorentz-violating corrections to the standard Aharonov-Bohm phase are very small, $\sim {\cal O}(\kappa^2)$, and hence undetectable by means of the current experimental techniques.

The non-topological phase of equation (\ref{9}) then becomes:
\begin{equation}
\delta \approx -\frac{2\pi \kappa_x eBR^2}{v_0}
\label{9a}
\end{equation}
From eq. (\ref{9}) it follows that the sensitivity to the Lorentz-violating parameter below the current upper bound $\kappa \lesssim 10^{-11}$ can be achieved with a rather strong magnetic flux in the solenoid. For instance, taking  $B=1$T, $R=0.5$cm,  and $v_0=0.6$ ($\sim 100$ keV electrons),  we find that the phase (\ref{9a}) will be detectable ($\delta \sim 1$rad) if $\kappa_x \gtrsim \cdot 2.5\cdot 10^{-12}$. However, in actual Aharonov-Bohm
 experiments the magnetic flux and inter-beam separation is much smaller, so that the sensitivity of order $\kappa_x \sim 10^{-7}$ can be achieved at best. This is related to the fact that the electron wavelength in typical interference experiments is very small ($\sim 3\cdot 10^{-10}$cm) and thus the whole interferometer, and  the size of the solenoid, has to be scaled down appropriately. Namely, the inter-beam distance must be of the order or less than the transverse coherence length of an electron wave packet   in order to observe the interference fringes produced by one electron at a time. 
 
However, since the phase shift in eq. (\ref{9}) is attributed to the local rather then topological interactions, there is no absolute necessity to follow the standard Aharonov-Bohm
 set-up. Instead we may place the solenoid outside the inter-beam  region (e.g., above the beam A) and assume it to be macroscopic.  Importantly,  we also require now that the beam $B$ propagates inside a Faraday cage which shields that beam from the induced electrostatic field from the solenoid, so that only beam $A$ experience its action. The phase shift now reads: 
\begin{equation}
\delta \approx  \frac{eBR^2}{v_0}\left[\kappa_y\log\left(\frac{l_2}{l_1}\right)+ \pi \kappa_x \right]~.  
\label{9a}
\end{equation}
Taking $\kappa_y=0$, or adjusting the position of the solenoid such that $l_1=l_2$,  
 we estimate a phase shift of 1 rad for $\kappa_x \approx 5\cdot 10^{-12}$ for the same solenoid as before. Thus, with 10\% (1\%) accuracy in phase measurements one order (two orders) of magnitude improvement of the currently available bounds  \cite{Wolf:2004gg} is expected. Similar sensitivity follows for the first contribution in (\ref{9a}),  for e.g. $l_2/l_1\sim 10$. One should point out here that both the sign and the magnitude of the phase shift depends on the position of the solenoid. This fact can be used in actual experiments to verify the Lorentz violating nature of the effect.  Sensitivity to the Lorentz violating effects of the  interference experiment described above can be further  improved e.g. by increasing magnetic flux in the solenoid. Note also that the effect crucially depends on the geometry of the magnetic source which, perhaps, can be favorably adjusted. 

\paragraph{Conclusion.}
In this paper we have suggested the study of possible violation of Lorentz invariance by particle interference experiments. A magnetic solenoid produces a static electric field which extends in the region outside of solenoid. Thus the solenoid in Lorentz-violating electrodynamics interacts locally with moving charged particles. As a result an extra path-dependent phase (\ref{9}) is generated.  

We have demonstrated that particle 
 interference experiments are sensitive to very small Lorentz-violating  effects and could provide important information on possible violation of Lorentz invariance in nature. Our estimates show that couple of orders of magnitude improvement of currently available bounds on $\vec\kappa$ will be available in such experiments.  On the theoretical side, it will be interesting to consider more general  (including Lorentz violation in matter sector) models of Lorentz-violating eletrodynamics in this context. 

\subparagraph{Acknowledgments.}   We would like to thank Alan Kostelecky for important comments which lead to a major revision of the preliminary version of this manuscript. We are also indebted to Tony Klein for useful discussions on particle interference experiments and Akira Tonomura for email correspondence on the experimental precision in phase shift measurements.  The work was supported by the Australian Research Council. 


\end{document}